\documentclass[submission, Proceedings]{SciPost}
\hypersetup{
    colorlinks,
    linkcolor={red!50!black},
    citecolor={blue!50!black},
    urlcolor={blue!80!black}
}

\usepackage[bitstream-charter]{mathdesign}
\DeclareSymbolFont{usualmathcal}{OMS}{cmsy}{m}{n}
\DeclareSymbolFontAlphabet{\mathcal}{usualmathcal}

\usepackage{siunitx}

\renewcommand{\eqref}[1]{(\ref{#1})}

\newcommand{\figref}[1]{Figure~\ref{#1}}
\newcommand{\tabref}[1]{Table~\ref{#1}}
\newcommand{\secref}[1]{Section~\ref{#1}}

\newcommand{\dn}{\ensuremath{d_{\rm n}}}
\newcommand{\pow}[2]{\ensuremath{#1\!\times\!10^{#2}}}
\newcommand{\ecm}{\ensuremath{\si{\elementarycharge}\!\cdot\! \rm cm}}
\newcommand{\magHg}{\ensuremath{{}^{199}\text{Hg}}}

\newcommand{\R}{\mathcal{R}}

\numberwithin{equation}{section}
\numberwithin{figure}{section}
\numberwithin{table}{section}

\begin{document}

\begin{center}{\Large \textbf{
The search for the neutron electric dipole moment  at PSI\\
}}\end{center}

\begin{center}
G.~Pignol\textsuperscript{1$\star$} and
P.~Schmidt-Wellenburg\textsuperscript{2$\star$}
on behalf of the nEDM collaboration
\end{center}

\begin{center}
{\bf 1} Universit\'{e} Grenoble Alpes, Centre National de la Recherche
Scientifique, Grenoble INP, LPSC-IN2P3, Grenoble, France
\\
{\bf 2} Paul Scherrer Institute, 5232 Villigen, Switzerland
\\
*philipp.schmidt-wellenburg@psi.ch
\end{center}

%=============================================
\begin{center}
\today
\end{center}

% For convenience during refereeing (optional),
% you can turn on line numbers by uncommenting the next line:
%\linenumbers

\definecolor{palegray}{gray}{0.95}
\begin{center}
\colorbox{palegray}{
  \begin{tabular}{rr}
  \begin{minipage}{0.05\textwidth}
    \includegraphics[width=24mm]{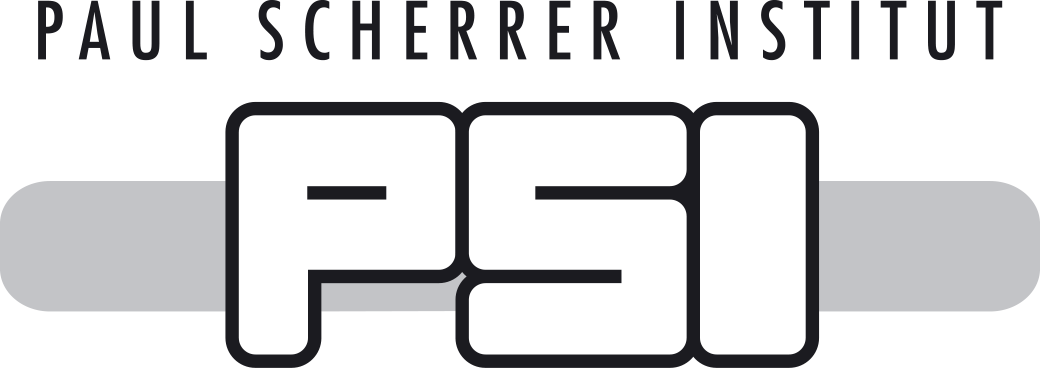}
  \end{minipage}
  &
  \begin{minipage}{0.82\textwidth}
    \begin{center}
    {\it Review of Particle Physics at PSI}\\
    \doi{10.21468/SciPostPhysProc.2}\\
    \end{center}
  \end{minipage}
\end{tabular}
}
\end{center}

%========================================

\section*{Abstract}
{\bf \boldmath
The existence of a nonzero permanent electric dipole moment (EDM) of
the neutron would reveal a new source of CP violation and shed light
on the origin of the matter--antimatter asymmetry of the Universe.
The sensitivity of current experiments using stored ultracold neutrons
(UCN) probes new physics beyond the TeV scale.  Using the UCN
source at the Paul Scherrer Institut, the nEDM collaboration has
performed the most sensitive measurement of the neutron EDM to date,
still compatible with zero ($|d_n|<1.8\times 10^{-26} \, e {\rm cm}$, C.L.\,90\%).
A new experiment designed to improve the sensitivity by an order of
magnitude, n2EDM, is currently in construction.  }

\setcounter{section}{27}
\label{sec:nEDM}

\subsection{Introduction}
\label{nEDM:sec:Intro}

\begin{figure}
	\centering
	\includegraphics[width=0.7\columnwidth]{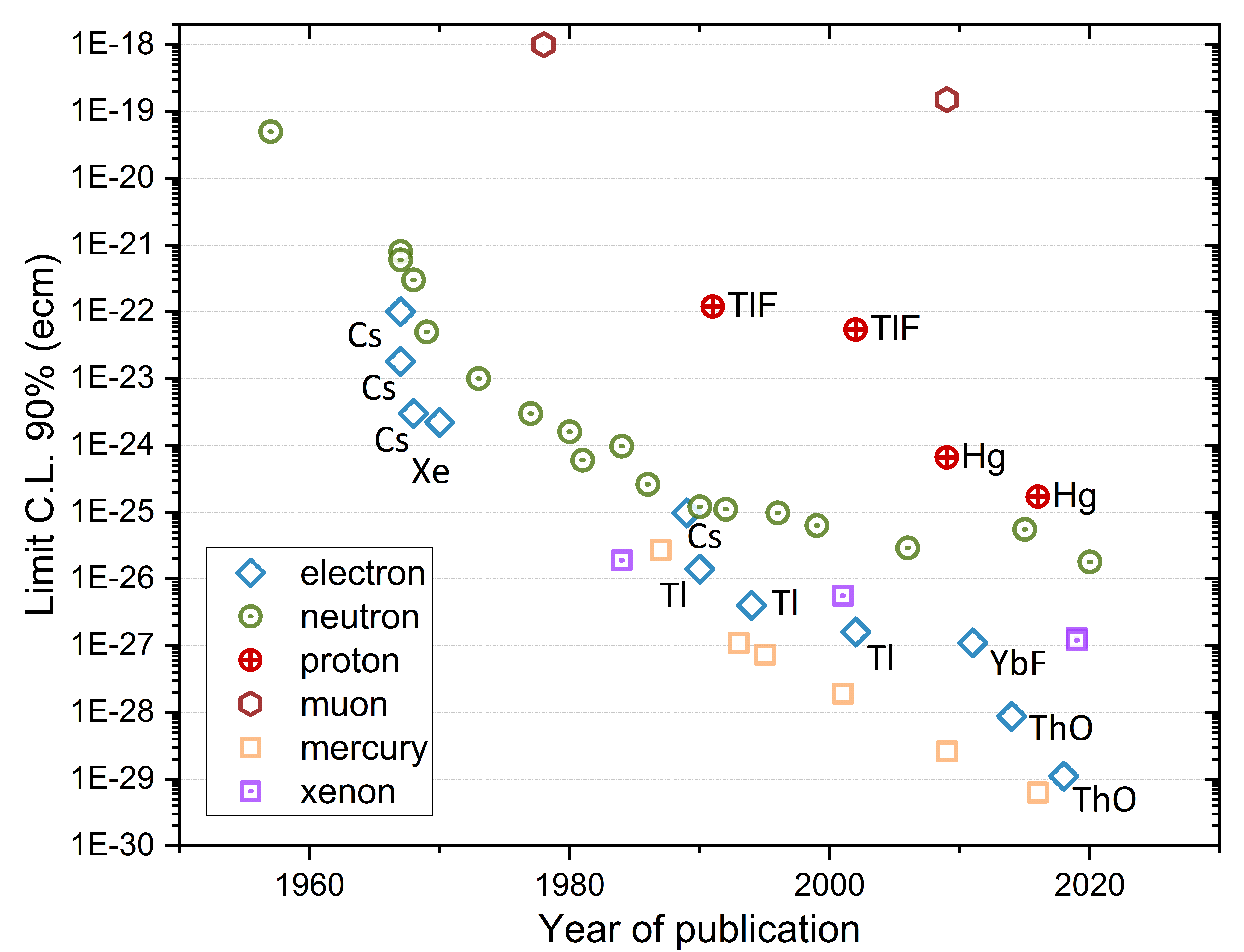}
	\caption[EDM limits]{History of upper limits (90 \% C.L.) for
          the EDM of various systems. Image first published in~\cite{Kirch:2020lbo}.}
	\label{nEDM:fig:history}
\end{figure}

The permanent electric dipole moment (EDM) $d$ of a simple quantum
system of spin 1/2 represents the coupling between the particle spin
and an externally applied electric field $\vec{E}$, in the same way
that the magnetic dipole moment $\mu$ quantifies the coupling between
the spin and an applied magnetic field $\vec{B}$.  The spin dynamics
is entirely described by the Hamiltonian
\begin{equation}
\label{nEDM:hamiltonian}
\hat{H} = -\mu \ \hat{\vec{\sigma}} \cdot \vec{B} - d \ \hat{\vec{\sigma}} \cdot \vec{E}, 
\end{equation}
where $\vec{\sigma}$ are the Pauli matrices.  Because
$\hat{\vec{\sigma}} \cdot \vec{E}$ is odd with respect to time
reversal, the CPT theorem implies that a non-zero EDM would result in
a violation of CP symmetry.  The search for a nonzero EDM was
initiated in the 1950's \cite{Smith1957}, applying the newly invented
resonance method with separated oscillating fields \cite{Ramsey1950PR}
on a thermal neutron beam. The quest for an EDM was then extended to
many other systems, as shown in \figref{nEDM:fig:history}, (see
\cite{Chupp2019RMP} for a review on EDM searches).  All experiments 
to date have reported results compatible with zero, despite the
million-fold improvement of the sensitivity of modern experiments.  As
discussed in the theory chapter of this volume, the present limits on
EDMs provide stringent constraints on theories beyond the Standard
Model of particle physics, which generally predict new sources of CP
violation and therefore non-zero EDMs.  The next generation of
experiments with improved sensitivity are motivated by the exciting
possibility of discovering a non-zero EDM induced by new physics at
the multi-TeV scale.

An international collaboration of 15 laboratories (\emph{the
  nEDM collaboration}) is conducting a long-term program at PSI to search for
the neutron EDM.  In 2009, the RAL/Sussex/ ILL
instrument\,\cite{Baker:2013iya}, which was previously used at the
Institut Laue Langevin in Grenoble for a long series of nEDM
measurements\,\cite{Pendlebury1984,Smith1990PLB,Harris:1999jx,Baker:2006ts},
was connected to the newly built high-intensity source of ultracold
neutrons\,\cite{Anghel2009,Lauss2014}.  After a phase of hardware
upgrades and commissioning of the instrument, data was
collected during 2015 and 2016.  This resulted in the
currently most precise measurement of the neutron EDM, $\dn =
\pow{(0.0\pm1.1_{\rm stat}\pm0.2_{\rm
    sys})}{-26}\,\ecm$\,\cite{Abel:2020gbr}.  This measurement, with
the \emph{single chamber instrument}, will be described in
\secref{nEDM:sec:Sec3}.  The construction of the new
\emph{double chamber instrument} (called n2EDM: the new neutron EDM
apparatus) started in 2018. It will be described in \secref{nEDM:sec:n2EDM}.
In the next section we elaborate on the main challenges to neutron EDM
searches.

\subsection{The three challenges for searches for the neutron EDM}
\label{nEDM:sec:Sec2}

The coupling in \eqref{nEDM:hamiltonian} leads to a precession of the
neutron spin around the fields at an angular frequency given by
$\omega =2\left(\mu B + d E\right)/\hbar$ in parallel electric and
magnetic fields.  In principle the EDM term can be separated from the
magnetic term by taking the difference of the frequency measured in
parallel and anti-parallel field configurations.  However, the electric term
that is to be measured is extremely small.  For $d=10^{-26} \, e
\rm{cm}$ and $E = 15 \, \rm{kV/cm}$, the spin would complete just
about two full turns per year, due to the electric term.  For the
detection of such a minuscule coupling, one needs (i) a long
interaction time with a large electric field, (ii) a high flux of
neutrons, and (iii) precise control of the magnetic field.  These
requirements constitute the three main challenges for the measurement.

\begin{figure}
  \centering
  \includegraphics[width=0.7\columnwidth]{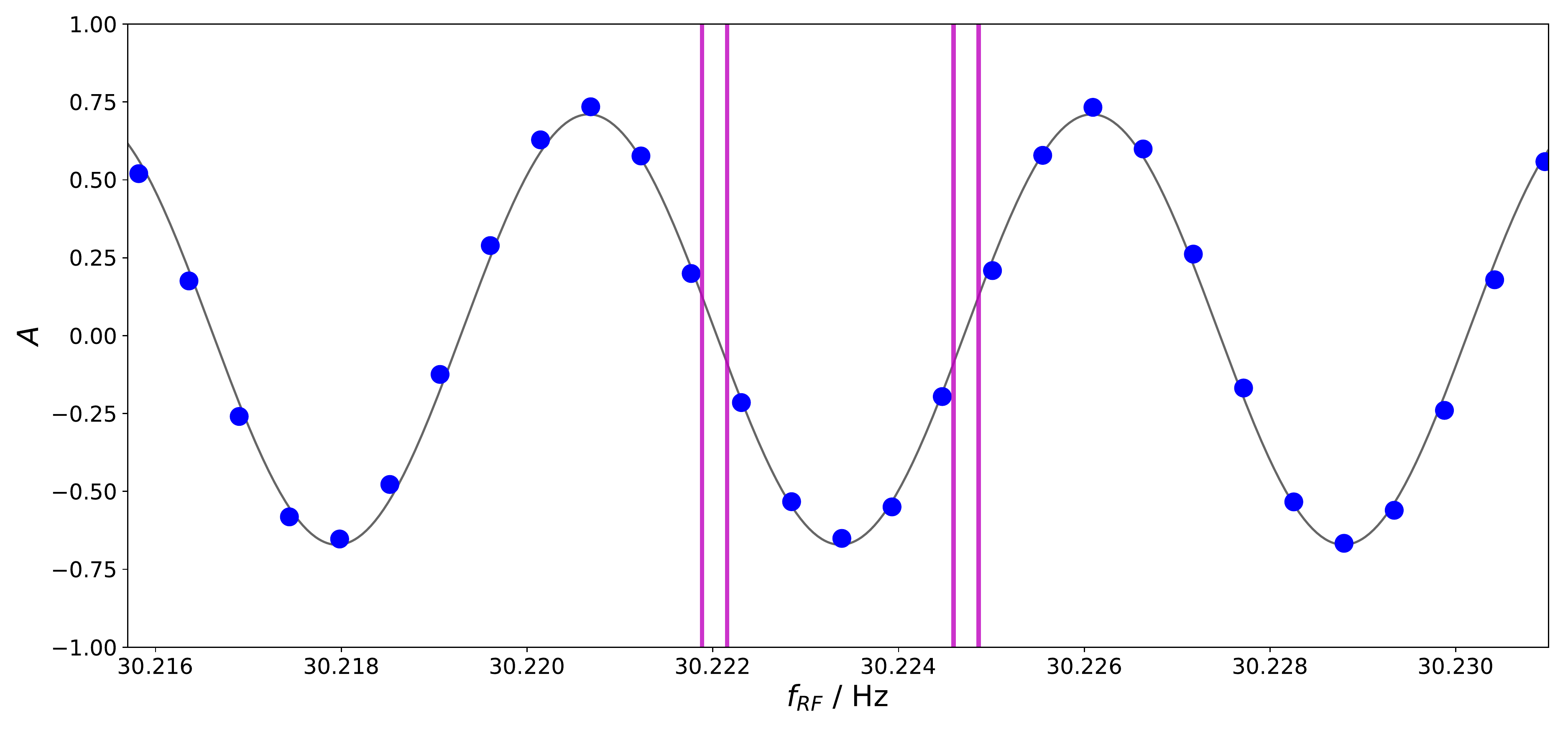}
  \caption[Ramsey fit] { Measurement of the asymmetry $A = (N_\uparrow
    - N_\downarrow)/(N_\uparrow + N_\downarrow)$ as a function of the
    applied frequency $f_{\rm RF}$ of the pulses.  Each point is a
    measurement cycle with a precession time of $T = \SI{180}{s} $
    performed with the single chamber apparatus in 2017.  The vertical
    bars show the position of the four ``working points'' used in the
    nEDM data-taking to maximize the sensitivity. The line is a fit of
    \eqref{nEDM:eq:CosineAppro} to the data.  }
  \label{nEDM:fig:RamseyFit}
\end{figure}

In many experiments, the neutron precession frequency is measured using Ramsey's
resonance method: neutrons with spins parallel to the magnetic field
are selected, then a first oscillating transverse magnetic-field pulse
is applied with a strength and duration adjusted to tilt the spin into
the plane transverse to the magnetic field.  The spins then precess
freely during a precession time $T$, after which a second pulse,
identical to and in phase with the first one, is applied.  At the end of
the process the neutron spins are analyzed in order to extract the
asymmetry $A$ of neutrons counted with spin up and down.  The
asymmetry is a function of the applied pulse frequency and of the
precession frequency to be measured, as shown in
\figref{nEDM:fig:RamseyFit}. By measuring the asymmetry, the
neutron precession frequency $f_n$ is extracted.  After combining several
measurements, aka cycles, of $f_n$ with different polarities of the
electric field the neutron EDM is measured with a statistical
sensitivity per cycle of
\begin{equation}
\sigma(d_n) = \frac{\hbar}{2 E T \alpha \sqrt{N}} 
\label{nEDM:Eq:stat_error_fn}, 
\end{equation}
where $N$ is the total number of neutron counts and $\alpha$ is the
visibility of the resonance, corresponding to the product of the neutron polarization at the end of the precession period and the analyzing power of the spin analyzer. It is apparent
from \eqref{nEDM:Eq:stat_error_fn} that the combination $ET$ enters
linearly in the statistical sensitivity and must be maximized (first
challenge) along with the statistical factor $\sqrt{N}$ (second
challenge).

The first neutron EDM experiments used beams of neutrons interacting
with the fields for only a few milliseconds.  The turning point for
higher sensitivities was the advent of ultracold neutron (UCN) sources
which permitted neutrons to be stored in a precession chamber for a
duration approaching the neutron half-life of 10 minutes.  Care must
be taken in the choice of materials constituting the precession
chamber in order to minimize neutron losses.  

In the single chamber apparatus at PSI,
the precession chamber was a cylinder of radius $23.5$~cm and height
$12$~cm, assembled from two aluminum electrodes coated with
diamond-like-carbon\,\cite{Atchison:2005you,Atchison:2006uas,
  Atchison:2006ps,Atchison:2007zza} and a polystyrene ring coated with
deuterated polystyrene\,\cite{Bodek2008}.  In average $N=15000$
neutrons per cycle were exposed to an electric field of $11$~kV/cm
during $T = 180$~s.

Based on experience and demonstrated developments, a double chamber apparatus was
designed.  Two vertically stacked chambers, with larger radii of
\SI{40}{cm} will sustain a larger electric field of opposite polarity
and store more neutrons.\newline \tabref{nEDM:TableStatSens} shows
the main parameters determining the statistical sensitivity.

\begin{table}[h]
\center
\begin{tabular}{c|c|c}
& single chamber (2016) & double chamber (projection) \\
\hline
\hline 
$N$ (per cycle) & 15'000 & 121'000\\
$T$   & 180 s & 180 s\\
$E$ & 11 kV/cm & 15 kV/cm\\
$\alpha$ & 0.75 & 0.8\\
\hline
\\[-1em]
%$\sigma(f_n)$ per cycle & \SI{9.6}{\micro Hz} & \SI{4.5}{\micro Hz}\\
%\hline
\\[-1em]
$\sigma(d_n)$ per day & 11 $\times$ 10$^{-26}$ \, $\ecm$ & 2.6 $\times$ 10$^{-26}$ \, $\ecm$
\\ \hline
\\[-1em]
%$\sigma(d_n)$ (final) & $9.5 \times 10^{-27}$ \, $\ecm$ & $1.1 \times 10^{-27}$ \, $\ecm$\\
%\hline
\hline
\end{tabular}
\caption[Experimental sensitivity]{ Comparison between (i)~the
  achieved performance of the single chamber apparatus during the
  datataking at PSI in 2016, (ii)~nominal parameters for the design of
  n2EDM.  }
\label{nEDM:TableStatSens}
\end{table}

The high statistical sensitivity must be combined with precise
control of the magnetic field: the third challenge.  This is
accomplished with a combination of magnetic shielding, the generation
of a stable and uniform magnetic field inside the shield, and
measurements of the magnetic field with atomic magnetometry.  In the
single chamber experiment, the change of the magnetic field between
reversals of the electric polarity (typically every 4 hours), needed
to be controlled at a level better than \SI{10}{fT}.
%\cymp{please explain what exactly is controlled to 10fT (done).} 
This was established by making sure that the Allan deviation for a field average over 4 hours was below \SI{10}{fT}.

For this purpose, the co-magnetometer technique\,\cite{Green1998,Ban2018} was
used.  Polarized \magHg{} atoms were injected in the chamber and the
precession frequency of the atoms was measured optically, providing
the magnetic-field average over the same time and almost the same
volume as the neutrons.

The mercury co-magnetometer is essential to control the residual time
variations of the magnetic field (both correlated and uncorrelated
with the electric polarity).  However, this comes at the price of
inducing a false EDM due to the combined effect of the relativistic
motional field $v \times E/c^2$ seen by the mercury atoms and the
magnetic field non-uniformities
\cite{Pendlebury:2004zz,Lamoreaux2005,Pignol2012PRA,Afach:2015ima}.  Due
to this important systematic effect, the control of the uniformity of
the magnetic field is of utmost importance. In particular,
ferromagnetic impurities close to the precession chamber(s) must be
avoided, and the residual large-scale magnetic gradients must be
minimized and measured with a combination of online and offline
methods.

\subsection{Measurement and result}
\label{nEDM:sec:Sec3}
The principal characteristic of the instrument operated between 2009
to 2017 at PSI was a single-chamber precession volume for UCN, which
at the same time contained spin-polarized \magHg{} atoms as
reference or cohabiting
magnetometer\,\cite{Green1998,Ban2018}.

\begin{figure}%
	\centering%
	\includegraphics[width=0.7\columnwidth]{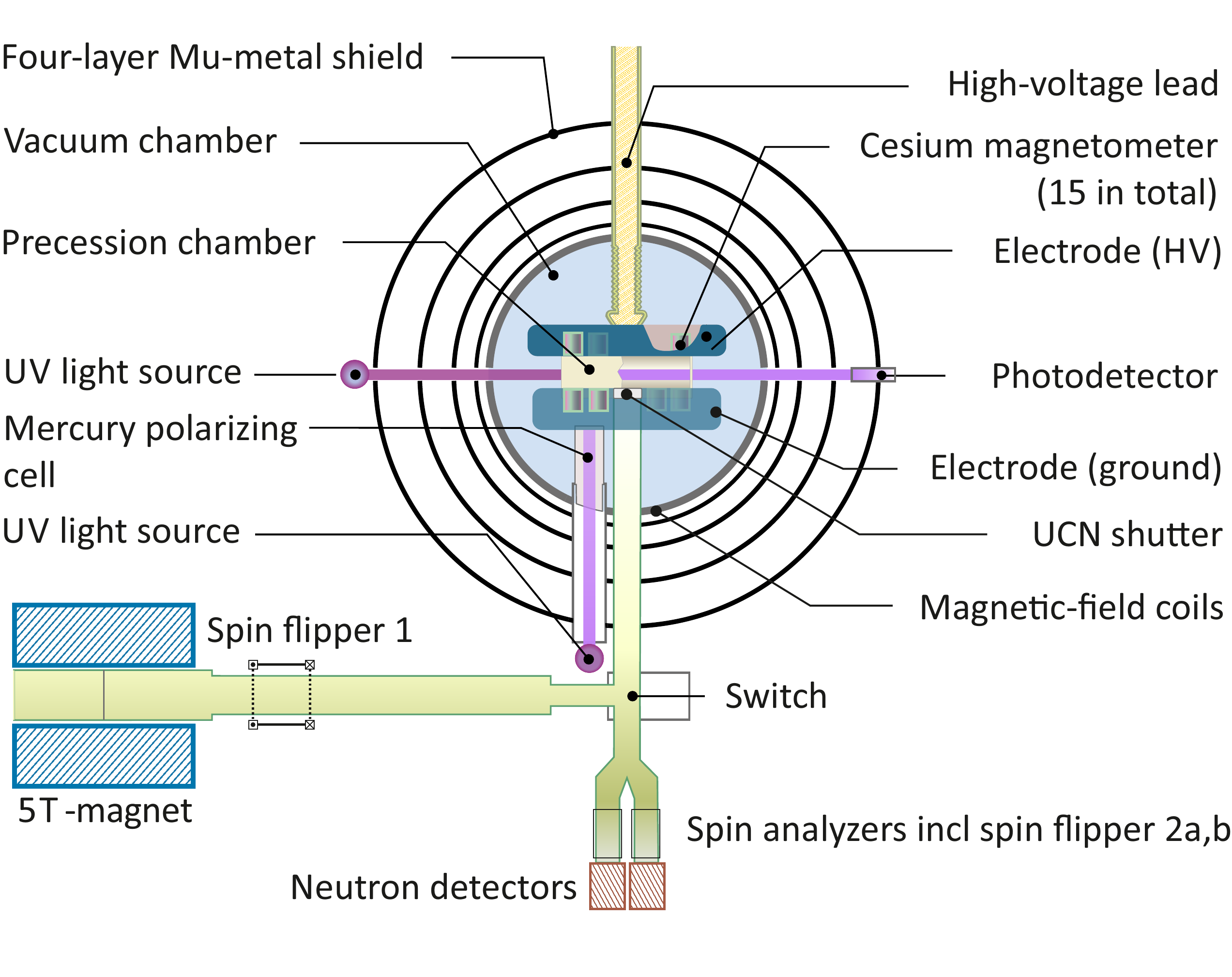}%
	\caption[Drawing of single chamber instrument]{Scheme of the
          single chamber experiment operated during 2009-2017 at
          PSI\@. Image first published in~\cite{Abel:2020gbr}.}%  
	\label{nEDM:fig:oILLSetup}%
\end{figure}

\figref{nEDM:fig:oILLSetup} shows a technical sketch of the
instrument. Ultracold neutrons from the PSI UCN
source\,\cite{Lauss2014,Bison2019EPJA} were polarized upon the passage
through the \SI{5}{\tesla} solenoid and entered the precession
chamber from the bottom.
The spin-manipulation and free precession of UCN and \magHg{} took
part here, \SI{125}{cm} above the horizontal beam line, inside a
4-layer mu-metal shield.  The top electrode made contact to the tip of
a high voltage~(HV) feed-through tested in operation up to
\SI{200}{kV}. An electric field of $E=\pm\SI{11}{kV/cm}$ was used for
data-taking.  The magnetic field, $B \approx \SI{1}{\micro\tesla}$,
was generated by a current of about \SI{17}{mA} in a $\cos\theta$-coil
wound directly onto the cylindrical vacuum tank.  In addition to the
$\cos\theta$-coil there were a total of 35 saddle and cylindrical
coils, aka trim coils, wound on the tank to adjust magnetic-field
gradients. Two of these saddle coils, on the top and bottom of the
vacuum tank, were used to set a small vertical magnetic-field gradient
$\partial B_z/\partial z$, for each sequence.  The
\magHg{}-comagnetometer measured the time and volume averaged magnetic
field within the precession chamber and was subject to the
above-described motional systematic effect. At the same time an array
of 15 optically-pumped Cs vapor
magnetometers~(CsM)\,\cite{Abel:2019cng}, mounted above and below the
chamber, was used to monitor the magnetic-field uniformity with a
sampling rate of \SI{1}{Hz}.  Another three coils, two of them in a
Helmholtz-like geometry and one a saddle coil, wound onto the outside
of the vacuum tank were used to generate the spin-manipulation pulses,
once the UCN and \magHg{}-atoms were inside
%\cymp{loaded? (changed to locked into)} 
the chamber, with
frequencies close to the resonance Larmor frequency of \magHg{} ($\sim
\SI{7.8}{Hz}$) and neutron ($\sim \SI{30.2}{Hz}$).

After the second $t=\SI{2}{s}$ long spin-flip pulse of the Ramsey
sequence the neutrons were counted in a spin-sensitive detection
system~\cite{Afach:2015hwe,Ban2016EPJA}. For each cycle,  from
%\cymp{sentence rearranged}
the recorded number of neutrons with spin up $N_u$ and down
$N_d$
the asymmetry
$A_i=\left(N_{u,i}-N_{d,i}\right)/\left(N_{u,i}+N_{d,i}\right)$ was computed. During data taking, the files containing the detector data were
blinded by injection of an artificial unknown EDM
signal~\cite{Ayres:2019uqu}, different for two distinct analysis
groups.

\begin{figure}
    \centering
    {\includegraphics[width=0.70\columnwidth]{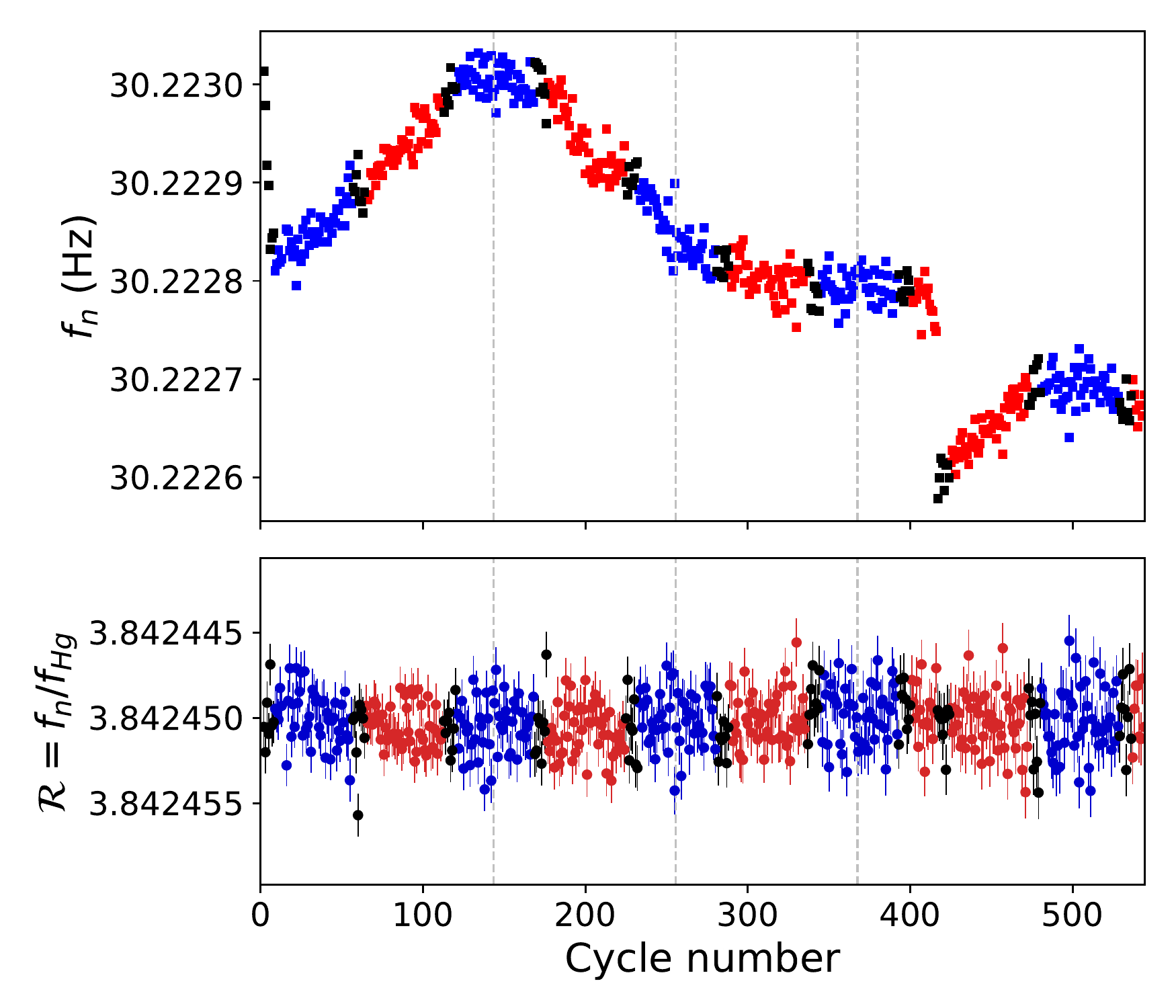}}
       \caption{ Plot of neutron frequency (top), $f_n$, and frequency
         ratio (bottom), $\R$, for a full sequence of nEDM data. Red
         data points indicate a positive voltage, while negative are
         marked blue. Black is used for cycles without electric
         field. A single EDM value is extracted for each sub-sequence,
         indicated by vertical dashed lines, before a weighted EDM
         average is calculated for the entire sequence. Figure reused from~\cite{Abel:2020gbr}.}
    \label{nEDM:fig:R-Sequences}
\end{figure}

\begin{figure}
    \centering
    \includegraphics[width=0.7\columnwidth]{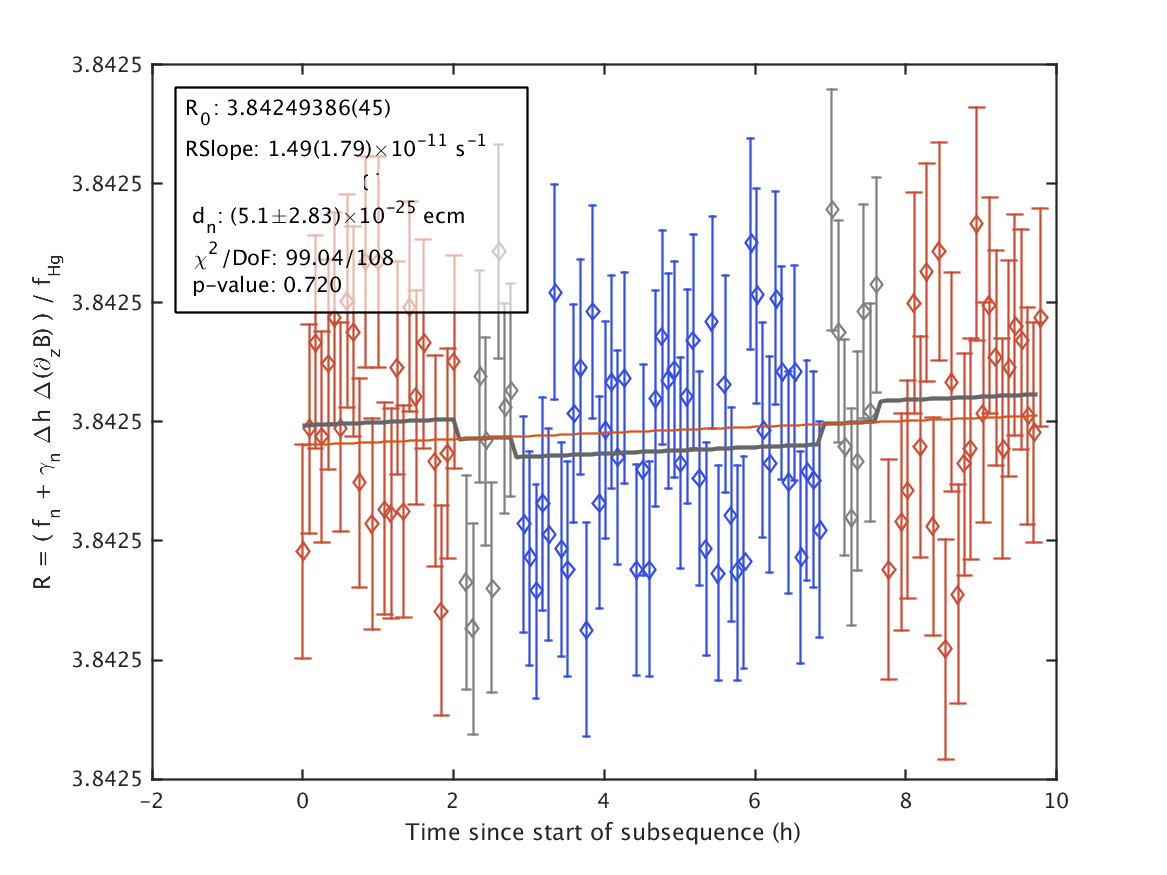}
    \caption{Subsequence with two polarity changes and a linear fit in time and $d_n E$ offsets. Note, that $E=-U/d$ hence positive electric fields (red) result from a negative charged electrode in \figref{nEDM:fig:R-Sequences}.}
    \label{nEDM:fig:Subsequence}
\end{figure}

\label{nEDM:sec:Result}
During the nEDM data acquisition period from July 2015 until December
2016 a total of 54\,068 cycles each with an average of about 11400
neutrons were recorded. The data were taken with different magnetic-field
configurations, e.g.\ $B$ up or downwards pointing with
$-\SI{25}{pT/cm}\geq\partial B_z/\partial z \leq \SI{25}{pT/cm}$. Each
of these sequences contained several hundred cycles and multiple
electric-field changes as can be seen in \figref{nEDM:fig:R-Sequences}. A
total of 99 sequences were analyzed.  In a first step, each sequence
was divided into sub-sequences including at least two changes of the
electric field polarity. The data of a sub-sequence, typically 114
cycles, was fit to

\begin{equation}
    A_i = A_{\rm off} \mp \alpha\cos\left(\frac{\pi f^{\prime}_{\rm rf}}{\nu}+\phi\right),
    \label{nEDM:eq:CosineAppro}
\end{equation}
where $f^{\prime}_{\rm rf}$ is the neutron spin flip frequency
corrected for magnetic-field drift using the measured $f_{\rm Hg}$ and
$\nu=1/(T+4t/\pi)$ is the width (FWHM) of the central fringe (see
\figref{nEDM:fig:RamseyFit}).  To extract the neutron resonance
frequency, $f_{{\rm n},i}$, the fit parameters
$A_{\rm off}$, $\alpha$ were fixed for each cycle and \eqref{nEDM:eq:CosineAppro} was solved for
$\phi = \pi f_{{\rm n},i}/\nu$.  \figref{nEDM:fig:R-Sequences}
bottom shows the ratio $\R_i=f_{{\rm n},i}/f_{{\rm Hg},i}$ for a full
measurement sequence.  An optimized analysis strategy was
implemented, accounting for all known effects \cite{Abel:2020gbr}
which affect the $\R$ ratio:
\begin{align}
\label{eq:AllShifts}
	\R = \left|\frac{\gamma_n}{\gamma_{\rm Hg}}\right| &\left(1+\delta_{\rm EDM}\right.  + \delta_{\rm EDM}^{\rm false}+\delta_{\rm quad} \\ \nonumber
							&\left. +\delta_{\rm grav} +\delta_{\rm T} +\delta_{\rm Earth}+\delta_{\rm light}+\delta_{\rm inc}+\delta_{\rm other} \right),  
\end{align}
in particular the EDM term $\delta_{\rm EDM} = 2E/(\hbar \gamma_n B)
d_n$.  In fact, the dominating effect is the gravitational shift
$\delta_{\rm grav} = G_{\rm grav} \langle z \rangle / B$, which is due
to the relative center-of-mass offset $\langle z
\rangle=-\SI{0.39(3)}{cm}$ between UCN and \magHg{}. This is both a
source of drifts (a nuisance) and also an excellent measure of the
effective vertical magnetic-field gradient $ G_{\rm grav}$.  In each
sub-sequence, the EDM signal $d_n^{\rm meas}$ and $\langle
\R\rangle$ are determined by fitting the $\R_i$ values, compensated for the relative
gradient drift, as a function of time and electric field by allowing,
also, for a linear time drift, as shown in
\figref{nEDM:fig:Subsequence}.
The measured $d_n^{\rm meas}$ for a given field configuration is
shifted by the term $\delta_{\rm EDM}^{\rm false} = 2E/(\hbar \gamma_n
B) d^{\rm false}$ corresponding to the motional false effect of
\magHg{} mentioned previous section. This effect depends on the
magnetic field gradients and can be expressed as~\cite{Abel:2019cng}:
\begin{equation}
\label{eq:falseEDM}
d^{\rm false} = \frac{\hbar}{8c^2}\left|\gamma_n \gamma_{\rm Hg} \right| R^2 \left(G_{\rm grav}+\hat{G}\right),
\end{equation} 
where $\hat{G}$ is the contribution from higher-order gradients and
does not produce a gravitational shift.  After correction of $\langle
\R\rangle$ and $d_n^{\rm meas}$ for $\delta_{\rm T}$ and $\delta_{\rm
  Earth}$, the contribution from $\hat{G}$, and minor systematic
shifts, the remaining shift is linear in $G_{\rm grav}$ and was
removed by a crossing point fit as shown in Figure~4
of~\cite{Abel:2020gbr}. The results of the crossing-point fit after
unblinding of the two analysis teams were
$d_{\times,1}\!=\!(-0.09\pm1.03)\!\times\!10^{-26}\,\ecm$,
$\R_{\times,1} = 3.8424546(34)$ with ${\rm \chi^2/dof}\!=\!106/97$ and
$d_{\times,2}\!=\!(0.15\pm1.07)\!\times\!10^{-26}\,\ecm$,
$\R_{\times,2} = 3.8424538(35)$ with ${\rm
  \chi^2/dof}\!=\!105/97$. The excellent agreement of both
$\R_{\times}$ values with each other and with the literature value
$\gamma_n/\gamma_{\rm Hg}=3.8424574(30)$~\cite{Afach:2015ima},
demonstrates the excellent control and understanding of all
magnetic-field-related shifts~\cite{Abel:2019cng}.

\subsection{n2EDM: The double chamber apparatus}
\label{nEDM:sec:n2EDM}
The concept and design of the new double chamber instrument,
n2EDM~\cite{Abel:2018yeo}, was based on maximizing the statistical
sensitivity of a single measurement, see \tabref{nEDM:TableStatSens},
while at the same time further reducing systematic effects.

\begin{figure}
    \centering
    \includegraphics[width=0.9\columnwidth]{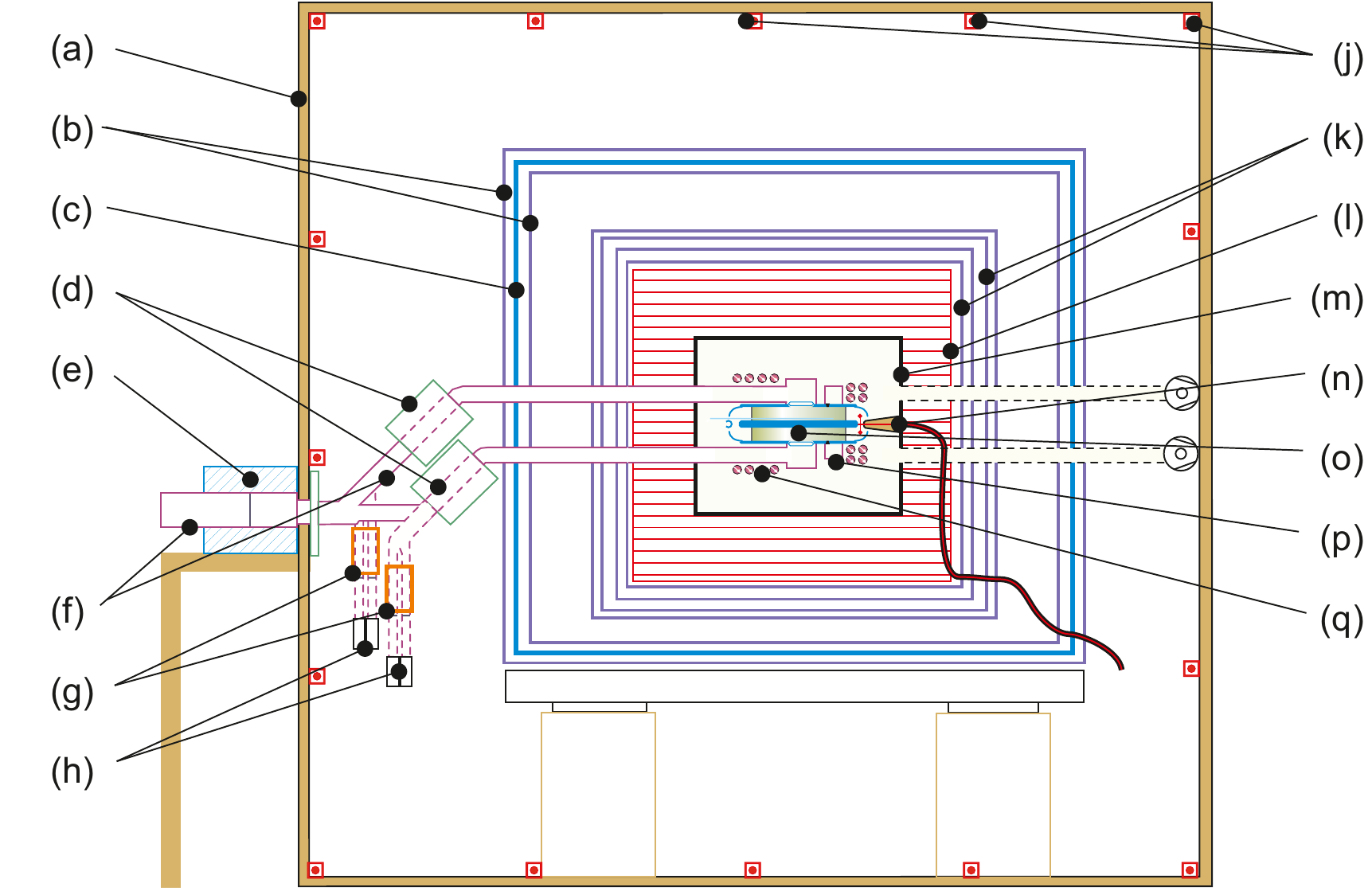}
    \caption{Sketch of the new double chamber instrument ``n2EDM'' at
      PSI from \cite{Abel:2018yeo}. (a)~Thermal shell, (b)~outer MSR shell, (c)~Eddy current shield, (d)~UCN switches, (e)~\SI{5}{T}-solenoid, (f)~UCN guides, (g)~fast adiabatic spin flippers, (h)~UCN detectors, (j)~AMS, (k)~inner MSR shells, (l)~magnetic field coils, (m)~vacuum chamber connected to turbo pumps, (n)~high voltage feed through and cable, (o)~double precession chamber with central electrode, (p)~\magHg polarization cell, (q)~cesium magnetometers. }
    \label{nEDM:fig:n2EDM}
\end{figure}

As can be seen in \figref{nEDM:fig:n2EDM}, 
%\cymp{add labels in \figref{nEDM:fig:n2EDM}} 
the new apparatus has two cylindrical storage chambers of diameter
$\varnothing \SI{80}{cm}$, made from proven materials, stacked one
above the other, separated only by a common high voltage electrode in
the center.  The UCN transport and storage layout was optimized for a
maximum number of neutrons per cycle using the established and bench
marked Monte Carlo code of the collaboration\,\cite{Zsigmond2018}.
This resulted in ultracold-neutron guides with constant effective
cross section and sub-nanometer roughness along the path up to the two
precession chambers which in turn are placed at the optimal height
relative to the beam line.

Both chambers are centered inside the same uniform magnetic field
generated by a main magnetic-field coil and an advanced trim-coil
system within a 6-layer magnetic and one-layer Eddy current
shield. First measurements of the quasi-static shielding factor in
2020 exceeded the specified value of $80\,000$ in all directions. This
is supplemented by an active magnetic shield~(AMS), similar to the active coil system used previously~\cite{Afach:2014kaa}, 
%\cymp{AMS  definition needed?} 
with eight degrees of freedom devised to
further improve the shielding factors at very low
frequencies. Dedicated coils were designed\,\cite{Rawlik2018} and
mounted onto the inner wall surfaces of the wooden thermal enclosure
to compensate gradient magnetic fields up to first order.  Hence,
neutrons and mercury inside the two precession chambers are exposed to
the same extremely low noise, highly uniform magnetic field while the
electric field points in opposite directions.  We expect that an
application of electric fields up to $|E|\geq\SI{15}{kV/cm}$ can be
achieved without difficulties, as the HV electrode is entirely
enclosed in a grounded Faraday cage.

All CsM are placed at ground potential and the previous limitation on the
electric-field strength due to flashovers along optical fibers of the
CsM can be ruled out.  The sensors were designed for an operation in
Bell-Bloom mode\,\cite{Grujic2015}, recording free spin-precession
waveforms for highest accuracy and with a sensitivity of better than
$200\,{\rm fT/\sqrt{Hz}}$.  This is an essential improvement for the accurate
determination of higher order magnetic-field terms relevant for the
correction of systematic effects.

Each precession chamber is connected via a UCN switch to a
simultaneous spin detection device featuring each two UCN detectors. A
gas mixture of CF4 and $^3{\rm He}$ is used for neutron detection. The
short scintillation pulse is registered by large surface
photo-multipliers and enables high count rate with very low background
counts from gamma rays or cosmic radiation.

In summary the new double chamber spectrometer, n2EDM, at PSI combines
the newest concepts and technologies while relying on proven
techniques and methods to improve the sensitivity frontier.

An attractive future option, which is described in great detail in
\cite{Pignol:2018cln}, eliminates the motional false EDM by
adjusting the magnetic-field strength so that the integral in
equation~(9) in \cite{Abel:2018yeo} vanishes. This magic field
configuration indicates a possible path to ultimate sensitivity using
the n2EDM spectrometer at PSI.

\subsection{Outlook and world-wide competition}
With the publication of the latest, most stringent limit of $d_{\rm
  n}<\SI{1.8e-26}{\ecm}$, PSI became the fourth member of the
exclusive club of institutes that have hosted a successful nEDM
search. It is now competing with a group of fierce and passionate
competitors from all around the
world\,\cite{Piegsa:2013vda,Picker2017,Ito:2017ywc,Ahmed:2019dhe,Wurm2019PPNS}
to break into the range of $\SI{1E-27}{\ecm}$ within the next
decade. A discovery of an nEDM or a further improved limit would
markedly and indelibly shape future models of particle physics
beyond the current Standard Model. \\

We wish to thank all the members of the nEDM collaboration for their dedication to the experimental effort, and the Paul Scherrer Institute for the fantastic hosting conditions.

\bibliography{nedm}

\nolinenumbers

\end{document}